\documentclass[runningheads]{llncs}
\usepackage{multirow}
\usepackage{graphicx}
\usepackage{svg}
\usepackage{todonotes}
\usepackage{braket}
\usepackage{amsfonts}
\usepackage{amsmath}
\usepackage{algorithm}
\usepackage{algpseudocode}
\usepackage{hyperref}

\begin{document}

\title{Quantum Annealers Chain Strengths: A Simple Heuristic to Set Them All}
%
%
\author{Valentin Gilbert\orcidID{0009-0001-1004-0204} \and
Stéphane Louise\orcidID{0000-0003-4604-6453}}
\authorrunning{V. Gilbert et al.}
%
\institute{Université Paris-Saclay, CEA-List, F-91120, Palaiseau, France \\
\email{\{valentin.gilbert, stephane.louise\}@cea.fr}}
\maketitle              
\begin{abstract}
    Quantum annealers (QA), such as D-Wave systems, become increasingly efficient and competitive at solving combinatorial optimization problems. However, solving problems that do not directly map the chip topology remains challenging for this type of quantum computer. The creation of logical qubits as sets of interconnected physical qubits overcomes limitations imposed by the sparsity of the chip at the expense of increasing the problem size and adding new parameters to optimize. This paper explores the advantages and drawbacks provided by the structure of the logical qubits and the impact of the rescaling of coupler strength on the minimum spectral gap of Ising models. We show that densely connected logical qubits require a lower chain strength to maintain the ferromagnetic coupling. We also analyze the optimal chain strength variations considering different minor embeddings of the same instance. This experimental study suggests that the chain strength can be optimized for each instance. We design a heuristic that optimizes the chain strength using a very low number of shots during the pre-processing step. This heuristic outperforms the default method used to initialize the chain strength on D-Wave systems, increasing the quality of the best solution by up to $17.2\%$ for tested instances on the max-cut problem.
\end{abstract}
%
%
%


\section{Introduction}

The idea of using Quantum Annealers to solve combinatorial problems is not new and was exposed by E. Farhi et al. \cite{farhi2001quantum}. Despite strong theoretical proofs based on the quantum adiabatic theorem, the speed-up brought up by quantum annealers still needs to be quantified for real useful application. It comes from the fact that quantum annealers implement a noisy version of the more general Adiabatic Quantum Computation (AQC) (the reader may refer to T. Albash et al. \cite{albash2018adiabatic} for an introduction to AQC). Indeed, D-Wave systems suffer from more than five different sources of Integrated Control Errors (ICE) \cite{DWaveInternalDocumentation}. The maximal number of qubits available on the quantum chip limits the size of the problem that can be solved. Minor embedding algorithms transform the initial problem into a new one that fits the sparsely connected quantum chip. The transformation consists of mapping the initial problem's logical variables to a set of physical variables that can be straightly encoded on the physical qubits of the quantum annealer. As mentioned in V. Gilbert et al. \cite{gilbert2023}, assessing the quality of an embedding is not trivial. The number of physical qubits used in the embedding can serve as a first quality indicator, but the physical qubit's topology used to encode the logical qubits can also be considered. To the authors' knowledge, a single contribution on this issue was made by E. Pelofske \cite{pelofske20234}. Topologies used for physical qubits are usually chains of qubits because this structure maximizes the number of potential connections of the logical qubit. E. Pelofske shows that the performance of the quantum solver is increased when logical qubits are encoded on cliques instead of chains. The minor embedding of a logical variable into a set of physical qubits requires the setting of an accurate ferromagnetic coupling. Current state-of-the-art methods often scan a static range of values to find the appropriate ferromagnetic coupler strength.

In this context, we explore the advantages and drawbacks of different logical qubit encodings. This first analysis shows that the minimum spectral gap varies depending on the qubit encoding, as well as the minimum chain strength required to maintain ferromagnetic couplings. We perform a detailed analysis of the average performance of the two main existing minor embedding methods. This analysis helps to select sets of instances on which we analyze the chain break tendency w.r.t the chain strength value. In particular, we discover that the optimal value of the chain strength varies depending on the embedding choice, which suggests that a per-instance chain strength setting method is desirable. The final contribution of this article is the design of a simple algorithm used to find appropriate values for the chain strength using very few pre-processing calls to the quantum computer. This new algorithm outperforms the default method implemented by D-Wave. 

The rest of the paper is organized as follows: Section \ref{section:2} introduces background about quantum annealing and minor-embedding methods. Section \ref{section:3} surveys the related work used in the study. Section \ref{section:4} describes the method and technical settings for the experiments. Section \ref{section:5} analyzes and proves the minimum spectral gap reduction induced by coupler strength rescaling. The Section \ref{section:6} gathers experiments on embedding and chain breaks that are used to build the algorithm presented in Section \ref{section:7}.

\section{Quantum Annealing and Minor Embedding Methods}\label{section:2}

D-Wave systems evolution is based on the transverse field Ising model. This model can be described with a linear interpolation of two Hamiltonians: a mixing Hamiltonian $H_\mathrm{M}$, which ground state (i.e., state of lower energy which can be degenerate) is easy to prepare, and a problem Hamiltonian $H_\mathrm{P}$, which ground state encodes the solution to the problem. The evolution of the system at annealing fraction $s = \frac{t}{T}$ is described by the Hamiltonian:

\begin{equation}
    H(s) = (1-s) H_\mathrm{M} + s H_\mathrm{P}
    \label{init_hamiltonian}
\end{equation}

where $T$ denotes the total run time of the quantum evolution and $t \in [0, T]$. The transverse field Hamiltonian $H_\mathrm{M} = \sum_{i=0}^{n-1} \sigma_i^x$ has its ground state defined by a uniform superposition of all the quantum states of the computational basis. The problem Hamiltonian can be fully specified by the user on a graph \(G_\mathrm{s}=(V_\mathrm{s}, E_\mathrm{s})\):
\begin{equation}
    H_\mathrm{P} = \sum_{v \in V_\mathrm{s}} h_v \sigma_v^z + \sum_{(u, v) \in E_\mathrm{s}} J_{uv} \sigma_u^z \sigma_v^z
\end{equation}

The ground state of the Hamiltonian $H_\mathrm{P}$ gives the solution to the Ising cost function minimization problem :

\begin{equation}
    \min C(\textbf{x}) = \sum_{v \in V_\mathrm{s}} h_v x_v + \sum_{(u, v) \in E_\mathrm{s}} J_{uv} x_ux_v 
    \label{ising_cost_function}
\end{equation}

where $x_u, x_v \in \{-1, +1\}$ and $h_v, J_{uv} \in \mathbb{R}$. The optimal solution is given by $\textbf{x*} = (x_0, x_1, ..., x_{n-1})$. The Hamiltonian $H(s)$, at a fixed annealing fraction $s$, corresponds to a Hermitian matrix $H_s$, and can be decomposed in terms of its eigenvalues $E_i(s)$ and eigenvectors $\ket{v_i(s)}$:

\begin{equation}
    H_s \ket{v_i(s)} = E_i(s) \ket{v_i(s)} \textrm{ with }E_0(s) < E_1(s) < ... < E_k(s)
    \label{eigen_equation}
\end{equation}

This decomposition is also called eigenenergies decomposition as the eigenvalues $E_i(s)$ correspond to the energy of each eigenvector $\ket{v_i(s)}$. The spectral gap of the Hamiltonian $\Delta_{\mathrm{min}}$ is the difference between the energy of the first excited state and the energy of the ground state at any annealing fraction $s$:

\begin{equation}
    \Delta_{\mathrm{min}} = \min_{0 \leq s \leq 1} ( E_1(s) - E_0(s) )
\end{equation}

The adiabatic theorem guarantees that a quantum state remains in its instantaneous ground state if $T$ is chosen large enough to smooth the quantum evolution. In the best case, the time $T$ scales as $O(1/\Delta_{\mathrm{min}}^2)$ \cite{albash2018adiabatic}.

When the Ising cost function of interest cannot be straigthly mapped on the chip topology, one has to use a method to minor embed the problem into the quantum chip. This problem is well defined by the theory of graph minors developed by Robertson and Seymour \cite{Robertson1995}. The problem is defined as follows: \\

\textit{Given a source graph \(G_\mathrm{s}=(V_\mathrm{s}, E_\mathrm{s})\), a target graph \(G_\mathrm{t}=(V_\mathrm{t},E_\mathrm{t})\), the aim is to find a mapping function \(\phi : V_\mathrm{s} \xrightarrow{} \mathcal{P}(V_\mathrm{t})\) such that : 
\begin{enumerate}
    \item each vertex \(v \in V_\mathrm{s}\) is mapped onto a connected subgraph \(\phi(v)\) of \(G_\mathrm{t}\).
    \item each connected subgraph must be vertex disjoint \(\phi(v) \cap \phi(v') = \emptyset\), with \(v \neq v'\).
    \item each edge \((u,v) \in E_\mathrm{s}\) is mapped onto at least one edge in \(E_\mathrm{t}\) : \(\forall (u,v) \in E_\mathrm{s}, \exists u' \in \phi(u), \exists v' \in \phi(v),~such~that~(u',v') \in E_\mathrm{t}\).
\end{enumerate}
}

For each vertex $v \in V_\mathrm{s}$, a ferromagnetic coupling strength $F_{\phi(v)}$ (also called chain strength) is applied to each edge of the subgraph $\phi(v)$. When this ferromagnetic coupling strength is the same for all ferromagnetic couplers, we note it $F_{\phi}$. In the rest of the paper, we refer to $V_s$ as the set of logical qubits and $V_t$ as the set of physical qubits. A chain break on a logical qubit $v$ means that at least one ferromagnetic coupling in $\phi(v)$ is corrupted. An edge of the subgraph $\phi(v)$ is a ferromagnetic coupling.

\section{Related Work}\label{section:3}
Two categories of methods are used to find a minor embedding of an Ising problem. The first category comprises a set of heuristics for which the source graph $G_s$ and target graph $G_t$ are given as input. The state-of-the-art implementation for this category is the CMR heuristic of Cai et al. \cite{cai2014practical}. The first step of this algorithm aims to find an initial embedding of each logical qubit with possible overlaps between their associated connected subgraph of physical qubits. The second step is a refinement that tries to reduce this overlap to return a valid embedding. The second category of methods considers that the source graph $G_s$ is a clique (i.e., a complete graph) and that the target graph $G_t$ is static. The state-of-the-art method used to generate Clique Minor Embeddings (CME) on D-Wave systems is an iterative method designed by Boothby et al. \cite{boothby2016fast}, which leverages the regular topology of the quantum annealer and considers inoperable qubits. Both CMR and CME methods have been extended with some pre or post-processing treatments to boost their performance (one example is Spring-based MinorMiner and Clique-Based MinorMiner \cite{Zbinden2020}).

Prior theoretical work by V. Choi \cite{choi2008minor} formulates two upper bounds on the minimum value of the ferromagnetic coupling strength $F_{\phi(v)}$. The first bound is straighty derived and corresponds to:

\begin{equation}
    F_{\phi(v)} < -\left( |h_v| + \sum_{u \in \mathrm{nbr}(v)} |J_{uv}|  \right)
    \label{eqn:choi_first_bound}
\end{equation}

where $\mathrm{nbr}(v)$ gives the set of nodes connected to $v$. This bound is fast to compute with $O(D)$ complexity, where $D$ is the vertex degree. V. Choi also details a more elaborated bound calculated in $O(DL)$ where $L$ is the chain length. In the paper of Fang et al. \cite{fang2020minimizing}, the authors derive a new tighter bound that can be computed in $O(D2^L)$. The main idea is to set the chain strength with a negative strength with its absolute value greater than the maximum potential energy gain of any configuration of the physical qubits that are not part of the ground state. Thus, this method scales exponentially w.r.t the size of the chain length. The authors of \cite{venturelli2015quantum} provided a second approach. They study the value of the optimal coupling strength by setting a global chain strength $F_{\phi}$. They observe that the optimal coupling value $F_\phi$ grows as the critical point of temperature of their embedded Sherrington Kirkpatrick model. Beyond this critical value, the success probability decreases. In the paper of Raymond et al. \cite{raymond2020improving}, the authors suggest that the chain strength should be tuned as:

\begin{equation}
    \lambda = \lambda_0 \sqrt{\sigma^2 N} = \lambda_0 \tau \sqrt{N}
\end{equation}

where $\sigma^2 = \frac{2}{N(N-1)} \sum_{(u, v) \in E_s}J_{uv}^2$ is the variance of the coupling strength (which is 1 for clique spin glasses). Their motivation is that a spin glass transition exists at optimal $\lambda_0$. However, $\lambda_0$ remains to be set empirically. This paper led to the default method \textit{uniform\_torque\_compensation} implemented on D-Wave systems \cite{DWaveInternalDocumentation}, which sets the value of the chain strength:

\begin{equation}
    F_\phi = - 1.414 \times \sqrt{\Bar{d}} \times \sqrt{ \frac{1}{|E_s|} \sum_{(u,v) \in E_s} J^2_{uv}}
    \label{eqn:torque_compensation}
\end{equation}

where $\Bar{d}$ is the average degree of the graph $G_s$. This formula comes from the fact that, for general Ising problem, the chain strength scales as $\tau \sqrt{N}$ where $\tau$ is the root mean square of the quadratic couplers.
In practice, the value of the chain strength is mostly set with a basic chain scan method used to maximize the average expectation value of the QA. For detailed experimental studies on chain strength scanning, the reader may refer to \cite{Hamerly2019,Willsch2022}. The chain scan method performs well but is very expensive in terms of the number of calls to the QA. The optimization of the chain strength is usually done based on the expectation value. A single advanced optimization method of the chain strength has been developed for the max-clique problem by H. Djidjev recently and is based on augmented lagrangian method \cite{djidjev2023logical}. Recent studies have benchmarked the chain break properties. In E. Grant et al. \cite{Grant2022}, the authors find that the chain break concentrates on specific locations of the D-Wave \textit{2000Q} processor and design post-processing strategies to limit these bias. In another recent paper of Pelofske \cite{pelofske2023comparing}, it is seen that the approximation ratio is inversely correlated with the rate of getting chain breaks. The author shows that the optimal chain strength is also conditioned by the density of the problem and the type of quantum solver used. We reuse these conclusions as well as the first theoretical bound given in equation \ref{eqn:choi_first_bound} to create an efficient method to set the chain strength.

\section{Method}\label{section:4}

This article solves two types of problems: the weighted Ising problem and the max-cut problem. Section \ref{section:5} is based on a non-degenerate instance of the weighted Ising problem. This problem consists in minimizing the Ising cost function formulated in equation \ref{ising_cost_function}. The minimum spectral gap of the problem $\Delta_\mathrm{min}$ is calculated using the exact diagonalization of the Hermitian matrix $H_s$ at each step of the annealing schedule $s \in [0, 1]$. The schedule used is the one corresponding to D-Wave \textit{Advantage6.4}. The detailed description of this schedule can be downloaded at \cite{DWaveInternalDocumentation}. 

The max-cut problem is used in Sections \ref{section:6} and \ref{section:7}. The weights $J_{uv}$ are set to 1 for all edges, $h_{v}$ weights are set to 0. All the experiments are run at a constant annealing time of $100 \mu s$. We use uniform logical weight spreading, majority vote to unembed the problem and the auto-scale method for weight rescaling. We do not use spin reversal methods or annealing offsets. The figures always show the chain strength in absolute value. The heatmaps from Fig \ref{embedding_comparison} are generated from sets of 100 instances for each size and density. Erdős-Rényi graphs and random d-regular graphs are generated using the Python \textit{networkx} library. The CMR heuristic (implemented in \cite{minorminer2023}) runs until a valid embedding is found. The CME heuristic is the default method provided by D-Wave \cite{DWaveInternalDocumentation}. For all the experiments, we select the first valid embedding found. The plots from Fig \ref{chain_break_comparison} are averaged over 30 instances of random Erdős-Rényi graphs of size $n=80$ and density $p=0.3$. The solver used is the \textit{Advantage6.4} and the number of shots is set to $1024$ for each chain strength. The plots from Fig \ref{chain_break_embedding} are obtained from a single Erdős-Rényi instance of size $n=60$ and density $p=0.4$. The solver used is the \textit{Advantage2\_prototype2.2} and the number of shots is set to $4096$ for each chain strength and for the runs with \textit{uniform\_torque\_compensation} method. The results obtained in Table \ref{chain_strength_setting_perf} are obtained with solvers \textit{Advantage6.4} and \textit{Advantage2\_prototype2.2}. The results are averaged over 30 instances of Erdős-Rényi graphs for each specified size and density. The number of shots in the preprocessing method is set to $128$. The final run used to retrieve the statistics is set to $4096$. The run with \textit{uniform\_torque\_compensation} method is also set to $4096$.

The metric used to compare the performance of heuristics that generate minor embeddings in Fig \ref{embedding_comparison} is defined as the average ratio of the number of qubits used by the CMR method and by the CME method:
\begin{equation}
    r_{emb} = \frac{1}{N_\mathrm{p}} \sum_{i}^{N_\mathrm{p}} \frac{n_{i}^{\mathrm{CMR}}}{n_i^{\mathrm{CME}}}
    \label{embedding_ratio}
\end{equation}

where $N_\mathrm{p}$ is the number of instances, $n_{i}^{\mathrm{CMR}}$ and $n_i^{\mathrm{CME}}$ is the number of physical qubits found by the embedding method CMR resp. CME on instance $i$. The average breaking chain rate $\epsilon_{b}$ of a single shot is defined by:
\begin{equation}
    \epsilon_{b} = \frac{1}{n_{lq}} \sum_{i=1}^{n_{lq}} p_b(i)
\end{equation}

with $n_{lq}$ the number of logical qubits. $p_b(i)$ is set to $1$ if the logical qubit contains at least one broken chain and $0$ otherwise. The average breaking chain rate of a serie of shots is given by:

\begin{equation}
    \bar{\epsilon_b} = \frac{1}{n_s} \sum_{i=1}^{n_s} \epsilon_{b}^{(i)}
    \label{eqn:break_rate}
\end{equation}

where $n_s$ defines the number of shots used in the experiment.

\section{Logical Qubit Structure} \label{section:5}

\begin{figure}[t!]
    \centering
    \includegraphics[page=1,width=\columnwidth]{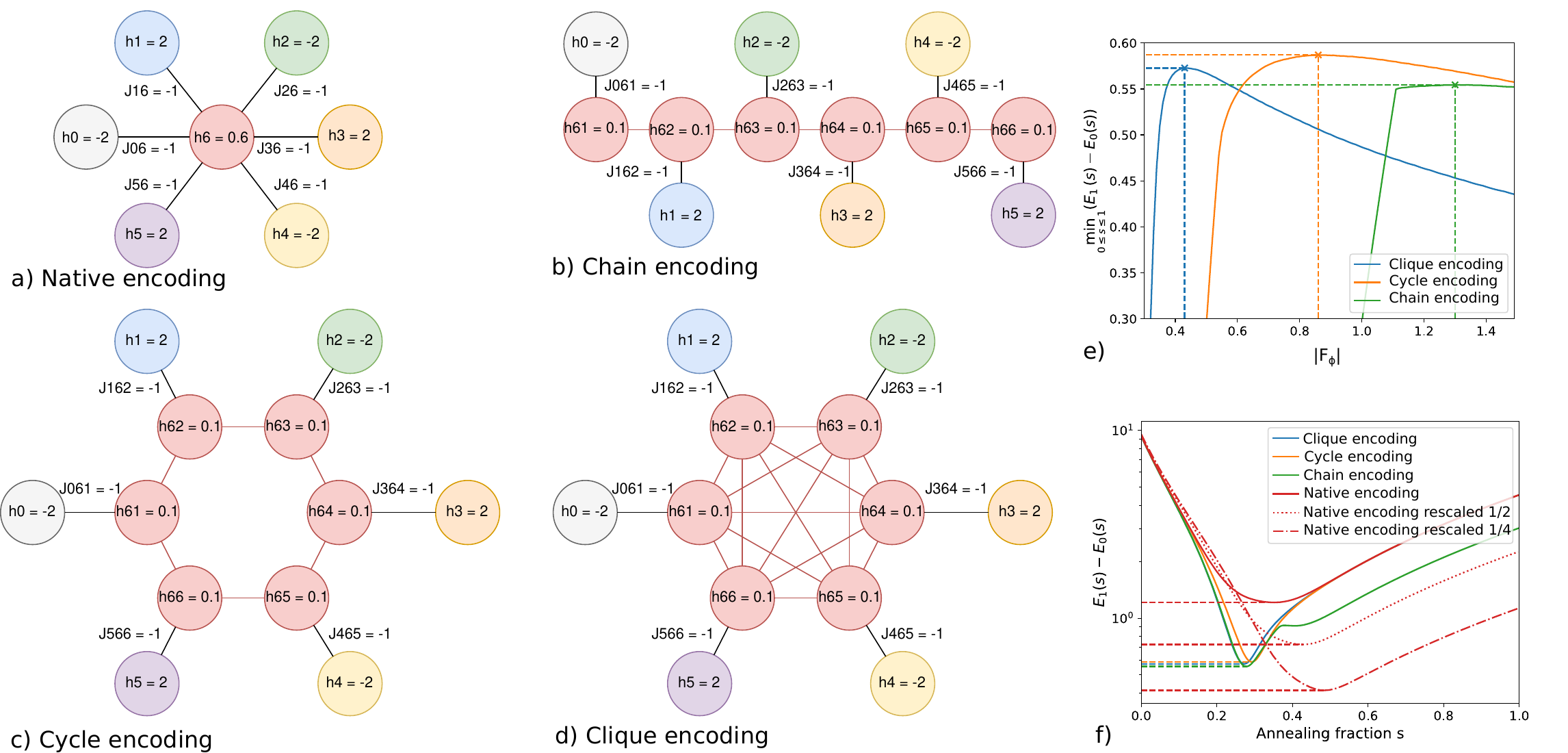}
    \caption[]{Minimum spectral gap evaluations considering different types of logical qubits encoding a) Native Ising problem instance b) Same instance as in a. with the red qubit encoded as a chain of physical qubits c) Same instance as in a. with the red qubit encoded as a cycle of physical qubits d) Same instance as in a. with the red qubit encoded with a clique of physical qubits. Black edges represent logical couplers and red edges represent ferromagnetic couplers parametrized by the chain strength. The auto-coupler strength $h_6$ is uniformly spread on each physical qubit. e) Evolution of $\Delta_\mathrm{min}$ considering the whole annealing schedule for different values of the global chain strength $|F_\phi|$. f) Spectral gap evolution of each encoding type programmed with the corresponding optimal chain strength indicated by dashed lines in e.}
    \label{fig_spectral_gap_log_qubit_structure}
\end{figure}

We study the evolution of the minimum spectral gap $\Delta_\mathrm{min}$ of a single max-cut instance by exhaustively simulating the quantum system evolution via matrix diagonalization. For this purpose, we design a single instance of an Ising problem with 6 nodes (see Fig. \ref{fig_spectral_gap_log_qubit_structure}.a). We choose three possible encodings for the physical qubits that replace the red logical qubit in \ref{fig_spectral_gap_log_qubit_structure}.a. The selected structures are: chain, cycle and clique. Each structure is respectively shown in Fig. \ref{fig_spectral_gap_log_qubit_structure}.b, \ref{fig_spectral_gap_log_qubit_structure}.c and \ref{fig_spectral_gap_log_qubit_structure}.d. Figure \ref{fig_spectral_gap_log_qubit_structure}.e shows the evolution of the minimum spectral gap of each encoding according to the chain strength value used to maintain the ferromagnetic coupling. For each encoding, the chain strength starts at the minimum analytical value, which is sufficient to maintain the ferromagnetic coupling of the logical qubit (0.3 for the clique, 0.45 for the cycle and 0.9 for the chain). Each embedding exhibits some sweet spots that maximize the minimum spectral gap. The size of the minimum spectral gap decreases when the chain becomes stronger. The optimal value of the chain strength decreases with the density of the logical qubit encoding. This is a desirable feature as D-Wave quantum processors have a limited working range for auto-coupler (\textit{h\_range}) and coupler (\textit{extended\_j\_range}) strengths. In addition, the maximum coupling range limit imposes that the total strength of the couplers linked to each qubit remains in a specific range. These two physical limitations lead to a global rescaling of coupler strengths when the values exceed these ranges.

Reusing the work of V. Choi \cite{choi2020effects}, it is straightforward to demonstrate that rescaling the coupler strength in the problem Hamiltonian also rescales with the same factor the minimum spectral gap of the Ising model Hamiltonian. Let the $\alpha$-rescaled Hamiltonian with $s_2 \in [0, 1]$ and $\alpha > 1$:
\begin{equation}
    H^{1/\alpha}(s_2) = (1-s_2)H_\mathrm{M} + s_2 \frac{1}{\alpha}H_\mathrm{P} 
\end{equation}

We take equation \ref{init_hamiltonian} as the initial Hamiltonian $H$ with annealing fraction $s_1 \in [0, 1]$. Consider the system:

\begin{equation}
    \begin{cases}
        \frac{H(s_1)}{1-s_1} = H_\mathrm{M} + \frac{s_1}{1-s_1}H_\mathrm{P} \\
        \frac{H^{1/\alpha}(s_2)}{1-s_2} = H_\mathrm{M} + \frac{s_2}{\alpha(1-s_2)}H_\mathrm{P}
    \end{cases}
\end{equation}

The equality $\frac{s_1}{1-s_1} = \frac{s_2}{\alpha(1-s_2)}$ can be solved with $s_1 = \frac{s_2}{(\alpha-1)(1-s_2)+1} $ and $s_2 = \frac{s_1}{\frac{1}{\alpha}(1-s_1)+s_1}$. Using this correspondance, we have:

\begin{equation}
    \frac{H^{1/\alpha}(s_2)}{1-s_2} = \frac{H(s_1)}{1-s_1}
\end{equation}

The rescaled Hamiltonian has the form:
\begin{equation}
    H^{1/\alpha}(s_2) = \frac{1-s_2}{1-s_1} H(s_1) = \left(1+ \left(\frac{1}{\alpha}-1 \right)s_2\right)H(s_1)
\end{equation}

The eigenenergies of the Hamiltonian are then rescaled with the same factor. According to equation \ref{eigen_equation}, we have:
\begin{equation}
    E^{1/\alpha}_i(s_2) = \left(1+ \left(\frac{1}{\alpha}-1 \right)s_2\right) E_i(s_1)
\end{equation}

Fig \ref{fig_spectral_gap_log_qubit_structure}.f. shows the effect of couplers strength rescaling on the spectral gap. This figure shows that rescaling the global coupling strength reduces the minimum spectral gap by the same factor. It also shifts the spectral gap to the right. Encoding logical qubits on a set of physical qubits has a detrimental effect on the spectral gap of the problem. The difference in spectral gap reduction between these encodings seems negligible compared to the rescaling of the weights. Logical qubit encodings that are dense, such as clique, require a lower coupling strength to maintain ferromagnetic coupling within the physical qubits. This type of encoding could be favored compared to chain encoding in specific cases to limit the effect of coupler strength rescaling.

\section{Embeddings and Chain Break Analysis}\label{section:6}


\begin{figure}[h!]
    \centering
    \includegraphics[page=1,width=\columnwidth]{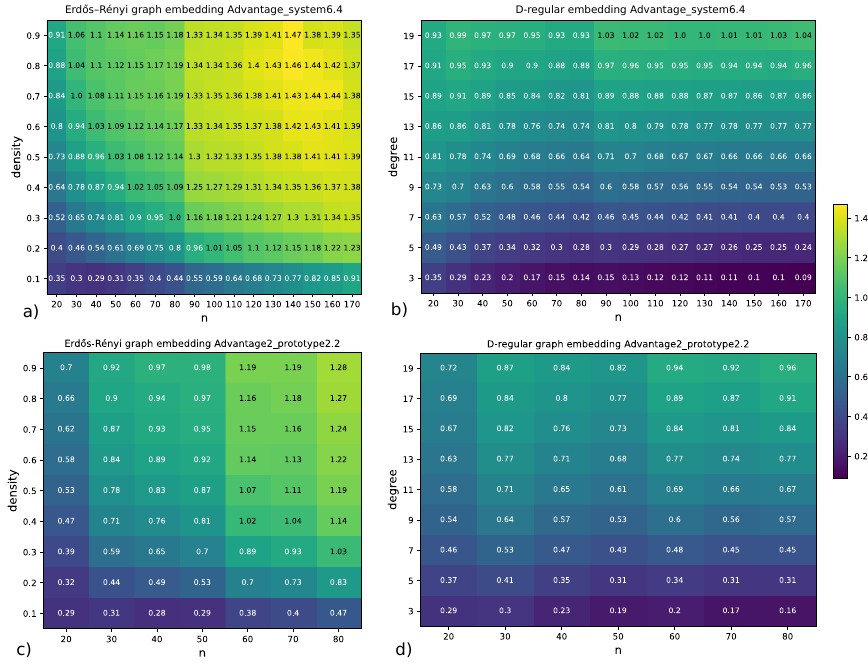}
    \caption{Heatmaps showing the average percentage overhead of the number of qubits used to embed similar instances using the CMR method compared to CME method. Each score is an average done over 100 instances. The score calculation in each cell is detailed in equation \ref{embedding_ratio}. a) and b) are embeddings generated for \textit{Advantage6.4} topology for Erdős-Rényi and d-regular graphs. b) and c) are embeddings generated for \textit{Advantage2\_prototype2.2} topology using the same instances. \textit{Advantage6.4} and \textit{Advantage2\_prototype2.2} can embed complete graphs of maximum size 82 and 174.}
    \label{embedding_comparison}
\end{figure}

\begin{figure}[t!]
    \centering
    \includegraphics[page=1,width=\columnwidth]{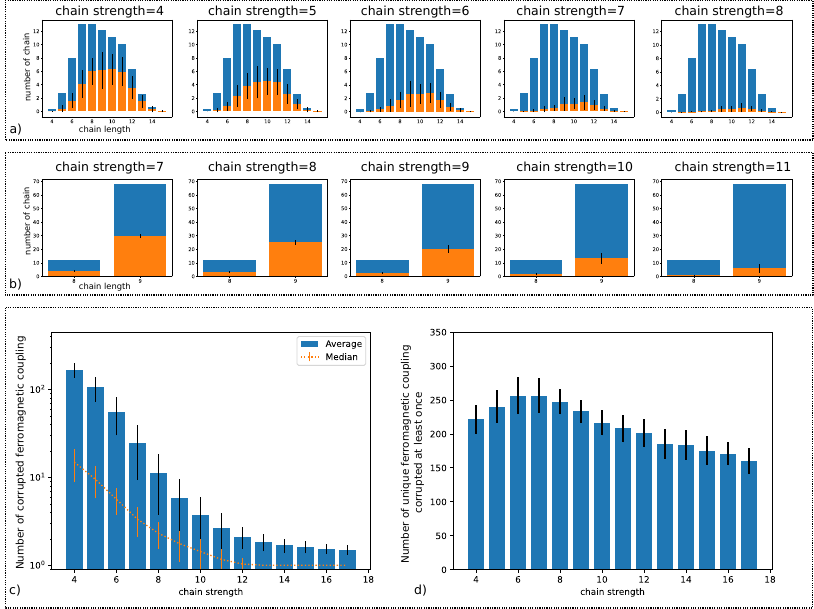}
    \caption{Statistics on the breaking chain rate (see equation \ref{eqn:break_rate}) averaged over 30 instances of Erdős-Rényi graphs of size $n=80$ and density $p=0.3$. a) resp. b) show the average chain length repartition of embedding obtained with CMR resp. CME methods. Blue bars show the average frequency of each chain length. Orange bars show the average breaking chain rate with a black error bar for the standard deviation. c) shows the average and median frequency of corrupted ferromagnetic couplings on CMR embeddings. d) shows the number of different ferromagnetic coupling that are corrupted at least once during the 1024 shots on CMR embeddings.}
    \label{chain_break_comparison}
\end{figure}

Figure \ref{embedding_comparison} compares the embedding performance in terms of the number of qubits used by the CMR  method \cite{cai2014practical} against the CME method \cite{boothby2016fast}. The comparison is done on Erdős–Rényi and d-regular graphs. It shows that CMR performs better on sparse graphs of small size, while CME is almost always preferred for large graphs. We use this heatmap to select sets of instances for which CMR and the CME methods produce embeddings with approximately the same number of physical qubits. Hence, we choose to generate 30 random instances of the max-cut problem with size $n=80$ and density $p=0.3$. The aim is to analyze the impact of the embedding method on the chain strength and length. CMR and CME produce embeddings with a similar number of physical qubits for these instances. To be fair in the comparison, we force the CMR method to generate minor embeddings with $\pm 1\%$ qubits compared to the embeddings generated by CME. Figure \ref{chain_break_comparison}.a and b. respectively show the repartition of the chain length for CMR resp. CME embeddings. While the CME method produces almost uniform chains of length 8 and 9, the CMR method produces chain lengths that approach a Gaussian distribution and vary from 4 to 15. The coupling strength required by the CME embeddings to obtain the same breaking chain rate is higher than for CMR embeddings. For the CMR method, the optimal solution probability occurs at chain strength 8, while this value is raised to 12 for CME embeddings. Figure \ref{chain_break_comparison}.d shows that the number of different corrupted couplings only slightly changes when the chain strength is increased. Fig \ref{chain_break_comparison}.c shows that when the chain strength is insufficient to maintain ferromagnetic couplings, the average number of corrupted ferromagnetic couplings is very high compared to the median. It suggests that the corruptions concentrate on the same few ferromagnetic couplings. When the chain strength value is increased, the average number of corruptions decreases. However, the number of different corrupted couplings remains the same, suggesting that the corruption becomes sparse when the chain strength is sufficient. We computed statistics on the number of corrupted ferromagnetic couplings per logical qubits. At the optimal chain strength value, the most likely scenario is to have only a single corruption of couplings on the logical qubit ($99.9\%$ cases) and very few chances of double corruption of couplings ($0.01\%$ cases).

\begin{figure}[b!]
    \centering
    \includegraphics[page=1,width=\columnwidth]{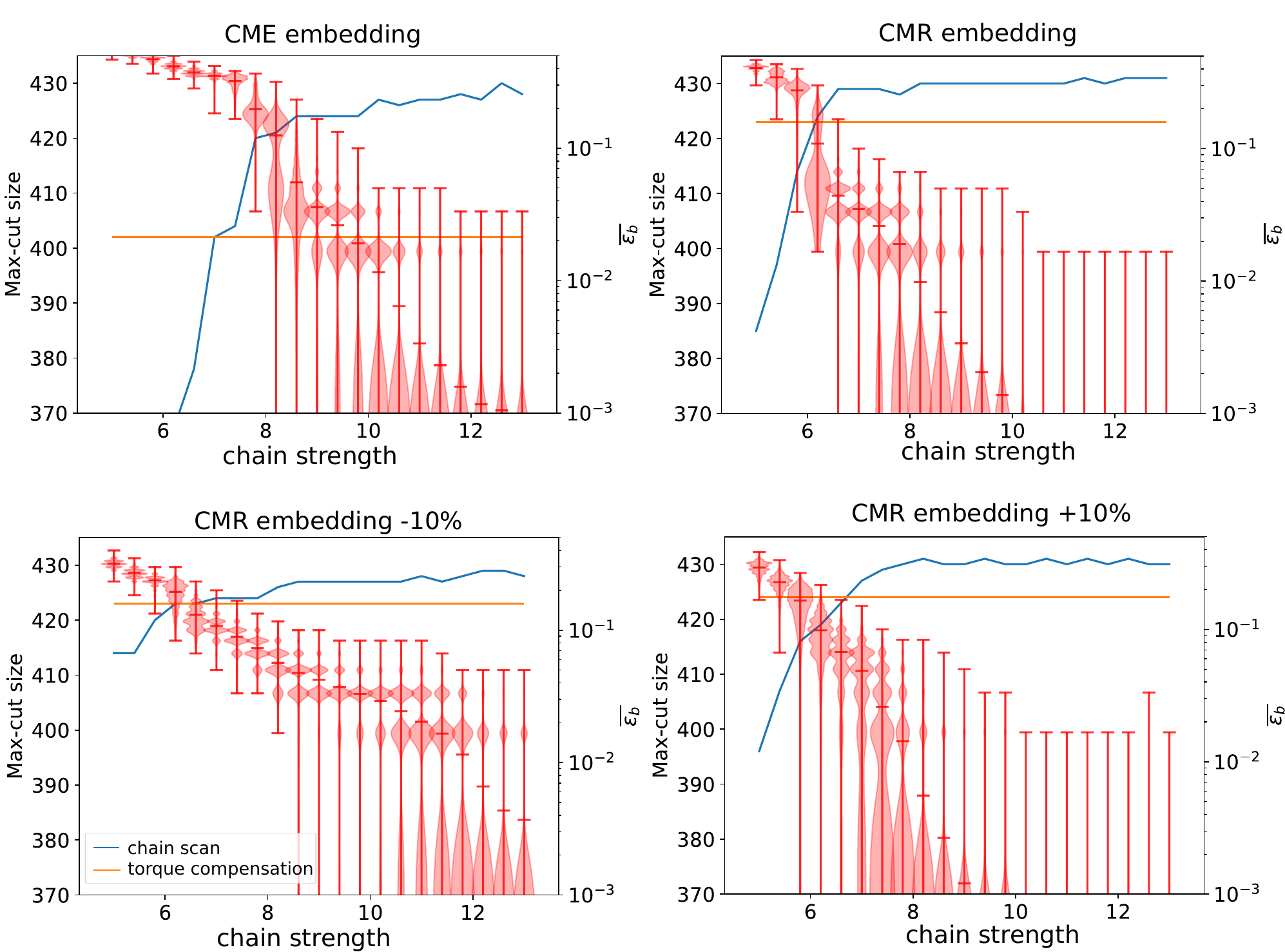}
    \caption{Maximum cut size obtained with a chain scan on four different embeddings of the same instance. The \textit{uniform\_torque\_compensation} is run once for each embedding (hence, it is independent of the chain scan). The CME and upper right CMR embedding have the same number of physical qubits ($\pm 1\%$). The two other embeddings have $10\%$ more and $10\%$ less qubits than the CME embedding. The red violins show the average breaking chain rate $\bar{\epsilon_b}$ related to the chain scan curve.}
    \label{chain_break_embedding}
\end{figure}

The above experiment suggests that the optimal chain strength value is related to the embedding. We select another instance that produces approximately the same number of qubits for both embedding methods for the \textit{Advantage2\_prototype2.2}. We generate four different embeddings for a single instance of the max-cut problem of 60 nodes and 0.4 density. The first embedding is generated by the CME method. We generate three other embeddings using the CMR method: one with a similar number of qubits as in the CME embedding ($\pm 1\%$), one which has $10\%$ less qubits than the CME embedding, one which has $10\%$ more qubits than the CME embedding. Figure \ref{chain_break_embedding} shows the best cut size found with each embedding. At first, we can see that the \textit{uniform\_torque\_compensation} method performs well in all the cases compared to the chain scan that requires heavy pre-processing ($93\%$ of the best cut size in the worst case with CME embedding). For each instance, the best cut size reaches a plateau when the chain strength becomes sufficient. This plateau is reached at different chain strength values for each instance. However, this plateau can also be located by only considering the average breaking chain rate from equation \ref{eqn:break_rate} (for example, considering that the chain break probability should remain under $2\times10^{-2}$). As specified in \cite{pelofske20234}, the chain break probability susceptible to provide the best result depends also on the quantum computer.

\section{Chain Strength Setting Heuristic}\label{section:7}

The previous section detailed our motivations for designing a heuristic to set the chain strength for each instance. Algorithm \ref{algorithm:chain_strength_search} describes this method in pseudo-code. The algorithm performs a binary search to find the optimal chain strength value within an initial interval of chain strengths set by the user $csInterval$. When the user does not specify it, the default interval's upper bound is set according to Choi's first bound (line 3):
\begin{equation}
    F_{\phi} = \min_{v \in G_t} {F_{\phi(v)}}
    \label{eqn:generalized_choi_first_bound}
\end{equation}

\begin{algorithm}
    \caption{Chain strength binary search}\label{alg:cap}
    \begin{algorithmic}[1]
    \Require $embInstance$, $cbInterval$, $csInterval$ (nullable)
    \Ensure $cbInterval[0] < cbInterval[1]$
    \State $hasConverged \gets False$
    \If{$csInterval$ is None}
        \State $csInterval \gets [0, getUpperBound(embInstance)]$ \Comment(see equation \ref{eqn:generalized_choi_first_bound})
    \EndIf

    \While{not $hasConverged$}
        \State $cs \gets csInterval[0] + (csInterval[0]-csInterval[1])/2$
        \State $res \gets runQA(embInstance, cs)$
        \State $\bar{\epsilon_b} \gets getChainBreakRate(res)$ \Comment(see equation \ref{eqn:break_rate})
        \If{$ cbInterval[0] \leq \bar{\epsilon_b} \leq cbInterval[1]$}
            \State $hasConverged \gets True$
        \ElsIf{$\bar{\epsilon_b} > cbInterval[1]$}
            \State $csInterval[0] \gets cs$
        \Else
            \State $csInterval[1] \gets cs$
        \EndIf
    \EndWhile
    \end{algorithmic}
    \label{algorithm:chain_strength_search}
\end{algorithm}

\begin{table*}[b!]
    \centering
    \caption{Performance comparison between D-Wave's default method (\textit{uniform\_torque\_compensation}) and our binary search heuristic. The column Best cut size shows the average of the maximum cut size obtained for each instance with the \textit{uniform\_torque\_compensation} method. The column Cut size improvement contains the minimum and maximum cut size improvement obtained with the chain strength found by Algorithm \ref{algorithm:chain_strength_search}. The standard deviation is shown is column std. The column Step counts the number of iterations required by our heuristic.}
    \begin{tabular}{|c|c|c|c|cccc|c|}
        \hline 
        \multicolumn{3}{|c|}{\textbf{Advantage2\_prototype2.2}} & Best cut size & \multicolumn{4}{c|}{Cut size improvement} & Step \\
        \hline
        Instance size & Density & Embedding & & min & max & mean & std & \\ 
        \hline
        \multirow{3}{*}{$n=40$} & 0.1 & CMR & 66.4 & $+0\%$ & $+0\%$ & $+0\%$ & $0\%$ & 5.4 \\ \cline{2-9}
         & 0.5 & CMR & 243 & $+0\%$ & $+2\%$ & $+0.2\%$ & $0\%$ & 5.8 \\ \cline{2-9}
         & 0.9 & CMR & 362.8 & $+2.1\%$ & $+8.2\%$ & $+5\%$ & $1.6\%$ & 4.6 \\ 
        \hline
        \hline
        \multirow{3}{*}{$n=80$} & 0.1 & CMR & 235.5 & $+0\%$ & $+0\%$ & $+0\%$ & $0\%$ & 4.7 \\ \cline{2-9}
         & 0.5 & CME & 804 & $+9.8\%$ & $+17.2\%$ & $+12.5\%$ & $0.2\%$ & 4.2 \\ \cline{2-9}
         & 0.9 & CME & 1435 & $+2\%$ & $+4.7\%$ & $+3.2\%$ & $0.6\%$ & 4.2 \\
         \hline
         \multicolumn{3}{|c|}{\textbf{Advantage6.4}} & Best Cut size & \multicolumn{4}{c|}{Cut size improvement} & Step \\
        \hline
        Instance size & Density & Embedding & & min & max & mean & std & \\ 
        \hline
        \multirow{3}{*}{$n=100$} & 0.1 & CMR & 355.9 & $+0\%$ & $+0.3\%$ & $+0\%$ & $0\%$ & 4.5 \\ \cline{2-9}
         & 0.5 & CME & 1271.4 & $+5.6\%$ & $+14.5\%$ & $+8.8\%$ & $1.8\%$ & 2.7 \\ \cline{2-9}
         & 0.9 & CME & 2243 & $+1.4\%$ & $+3.7\%$ & $+2.5\%$ & $0.5\%$ & 3.7  \\ 
        \hline
        \hline
        \multirow{3}{*}{$n=170$} & 0.1 & CMR & 950.8 & $-2.1\%$ & $+0.6\%$ & $-0.5\%$ & $0.5\%$ & 2.1  \\ \cline{2-9}
         & 0.5 & CME & 3631.4 & $+2.8\%$ & $+6.2\%$ & $+4.5\%$  & $0.7\%$ & 2.1 \\ \cline{2-9}
         & 0.9 & CME & 6519.4 & $+0.4\%$ & $+1.4\%$ & $+0.8\%$  & $0.2\%$ & 3.2 \\
         \hline
    \end{tabular}
    \label{chain_strength_setting_perf}
\end{table*}

At each iteration, the chain strength $cs$ is chosen as the midpoint of the chain strength interval $csInterval$ (line 6). The embedded instance $embInstance$ is then sent to the quantum annealer with the specified chain strength $cs$ (line 7). We then compute the breaking chain rate of the result according to equation \ref{eqn:break_rate} and check if the new breaking chain rate is in the interval specified by the user $cbInterval$. If this is the case, the algorithm converges (lines 9 and 10), and the loop breaks. If the chain break is higher than the upper bound of $cbInterval$, the chain strength is insufficient and requires an increase: we then rescale the lower bound of $csInterval$ (lines 11 and 12). In the other case, the upper bound of the chain strength interval is rescaled (line 14). The subtlety of this algorithm relies in the setting of the chain break interval $cbInterval$ which stops the algorithm when the sampling produces a breaking chain rate in this interval. This interval depends on the effective noise of the quantum computer and may also depend on the size of the instance. We test our heuristic and compare the results obtained with the default method implemented by D-Wave that relies on equation \ref{eqn:torque_compensation}. The comparison is done both on \textit{Advantage2\_prototype2.2} on small Erdős-Rényi instances ($n=40$ and $n=80$) and \textit{Advantage6.4} for large instances ($n=100$ and $n=170$) with density $p \in \{0.1, 0.5, 0.9\}$. For each instance set, we choose the embedding heuristic that provides the best average performance based on the size and density of the instance (see Fig \ref{embedding_comparison}). We empirically set the interval $cbInterval$ to $[6\times10^{-3}, 2\times10^{-2}]$ for the \textit{Advantage2\_prototype2.2} and to $[2 \times 10^{-2}, 5 \times 10^{-2}]$ for the \textit{Advantage6.4}. This range stays untouched during the whole experiment. Each round of pre-processing step is done with 128 shots. The final run of our heuristic, as well as the default D-Wave method, is programmed with 4096 shots for instances run on \textit{Advantage2\_prototype2.2} and 3072 shots for instances run on \textit{Advantage6.4}. Table \ref{chain_strength_setting_perf} shows the result of this experiment. With a global setting of the value for $cbInterval$, our heuristic is able to outperform in almost every case the default \textit{uniform\_torque\_compensation} method. The breaking chain rate can be obtained with relatively high fidelity in a few shots. It gives a considerable advantage to our method compared to the basic chain strength scans that always use the expectation value to find optimal values of the chain strength and, hence, require a very high number of shots. Our optimization method does, on average, between 2 and 6 iterations to find a suitable value for the chain strength (i.e., $\approx 750$ extra runs in worst cases). This running overhead is almost negligible compared to the 4096 and 3072 shots used to evaluate the final chain strength and the \textit{uniform\_torque\_compensation} method. The default \textit{uniform\_torque\_compensation} performs well on embeddings generated by CME for dense and sparse graphs that use CMR heuristic. Our heuristic seems to perform the best with mid-density instances by increasing cut size up to $17.2\%$ for the \textit{Advantage2\_prototype2.2} and $14.5\%$ for the \textit{Advantage6.4}.

\vspace{-2mm}
\section{Conclusion}

This paper presents a detailed analysis of the minimum spectral gap evolution considering different topologies for the set of physical qubits representing a single logical qubit. Each encoding requires a specific chain strength value to maintain ferromagnetic couplings between the physical qubits. The denser the logical qubit encoding is, the lower the chain strength has to be. This feature is desirable as the coupler strength rescaling, usually driven by chain strength values in combinatorial problems such as Max-cut, reduces the minimum spectral gap of the problem. This consideration could be included in future embedding heuristics designs to enhance the quality of the generated minor embeddings.

The analysis of the breaking chain rate considering different embeddings of the same instance has shown that the optimal chain strength varies depending on the embedding used. This experiment led to the design of a simple but fast heuristic used to optimize the chain strength for each instance. The heuristic converged in most cases in less than 5 pre-processing steps. This number has to be considered cautiously as it strongly depends on the chain break interval, which acts as the breaking condition of the heuristic. We used large intervals in this experiment. The precise identification of the breaking chain rate that produces the best results for each quantum computer could help to refine these bounds. Even if this new method does not provide better solutions than the basic chain scan, the original use of the breaking chain rate in the optimization process drastically reduces the pre-processing time required by a simple chain scan and does not require any assumption over the scanning range of chain strengths.

The performance of our heuristic can be questioned due to the relatively low gain on the max-cut size ($17\%$ in the best case). However, the default \textit{uniform\_torque\_compensation} method performed quite well compared to the chain scan method on the max-cut problem, meaning that this default heuristic is well-calibrated for this problem. However, even a small percentage gain in the cut size can reduce the Time To Solution of several orders of magnitude. We could not run such an experiment due to our restricted access time to the D-Wave quantum systems.

A detailed study on the determination of refined breaking chain bounds for each solver is a relevant perspective. This heuristic could also be used as a fast pre-processing routine to find relatively good global chain strengths, followed by a refinement step optimizing each ferromagnetic coupling strength.

\bibliographystyle{splncs04}
\bibliography{bibliography}

\end{document}